\documentclass[aps,12pt,nofootinbib,preprintnumbers]{revtex4-1}
\usepackage{amstext}
\usepackage[T1]{fontenc}
\usepackage[utf8]{inputenc}
\usepackage{amssymb,amsmath}
\usepackage{graphicx}
\usepackage{enumerate}

\begin{document}

\title{Cumulant Expansion in Gluon Saturation, and Five and Six-Gluon Azimuthal
Correlations}

\author{\c{S}ener \"{O}z\"{o}nder}
\email{ozonder@uw.edu}
\affiliation{Department of Physics, Istanbul Technical University, 34469 Istanbul, Turkey,}
\affiliation{Department of Physics, Koc University, 34450 Istanbul, Turkey}
\date{\today}

\begin{abstract}
Correlations between the momenta of the final state hadrons measured in proton or
nucleus collisions contain information that sheds light on the initial
conditions and evolutionary dynamics of the collision system. These
correlation measurements have revealed the long-range rapidity correlations
in p-p and p-Pb systems, and they have also made it possible to extract
the elliptic flow coefficient from hadron correlation measurements.
In this work, we calculate five- and six-gluon correlation functions
in the framework of saturation physics by using superdiagrams. We
also derive the cumulant expansion of the gluon correlators that is
valid in the gluon saturation limit. 
We show that the cumulant expansion
of the gluon correlators that is used for counting the number of diagrams to be calculated does not
follow the standard cumulant expansion. We also explain how these
findings can be used in obtaining experimentally relevant observables
such as flow coefficients calculated from correlations as well as
ratios of the correlation functions of different orders.
\end{abstract}

\maketitle

\section{Introduction}

The measured correlations between the final state hadrons in collisions
involving protons and nuclei contain information about the dynamics
of the collision and evolution of the produced particles \cite{DiFrancesco:2016srj}.
In nucleus-nucleus (A-A) collisions, it has been observed that the
particle pairs with the azimuthal angles $\phi_{1}$ and $\phi_{2}$
are maximally correlated when $\phi_{1}-\phi_{2}\sim0$ (collimation)
and $\phi_{1}-\phi_{2}\sim\pi$ (anticollimation) \cite{Wang:2013jro}.
Also, this correlation is maintained even if the pair is separated
by several units of rapidity. Such long-range azimuthal correlations
in A-A collisions have been ascribed to the collective radial flow of the quark-gluon plasma \cite{Dumitru:2008wn}. Radial flow gives rise to collectivity
in the hadronic spectrum where the momenta of the detected hadrons are
not random but correlated.

In small systems such as p-p and p-A, the collectivity in the produced
hadrons had not been observed in experiments or Monte Carlo simulations.
Also, on the theory side there was no such expectation of collectivity
in p-p and p-A collisions since it had been thought that fluid behavior
would not emerge in such small collision systems. However, the two-particle
correlation measurements in p-p collisions at $\sqrt{s}=7\,\mathrm{GeV}$
at the LHC revealed for the first time the existence of collimation and
anticollimation effects, which are long ranged in rapidity, appearing
at high-multiplicity events the so-called \textit{double ridge} \cite{Khachatryan:2010gv,Velicanu:2011zz,Li:2012hc,Wozniak:2017lgj}.
Later, the same ridge has also been seen in p-A collisions \cite{CMS:2012qk,Abelev:2012ola,Aad:2012gla,Milano:2014qua,ABELEV:2013wsa,Abelev:2014mda,Abelev:2014mva,Aad:2014lta}.
Afterwards, we predicted the existence
of higher-dimensional ridges that would appear in three-, four- and higher-dimensional
particle correlation functions \cite{Ozonder:2014sra}.

The observation of collectivity in p-p and p-Pb collisions later sparked an
interest in applying hydrodynamics to such small systems \cite{Bozek:2011if,Bozek:2012gr,Bozek:2013ska,Werner:2013ipa,Kozlov:2014fqa,Bzdak:2014dia}.
An alternative program that we pursue in this work, however, does
not assume hydrodynamical evolution. Instead, our approach here tracks
the origin of the collectivity in small systems to gluon saturation
in the target and projectile in p-p or p-A collisions
\cite{Dusling:2009ni,Dusling:2009ar,Kovner:1995ts,Kovchegov:1997ke,Gelis:2009wh,McLerran:1993ni,McLerran:1993ka,McLerran:1994vd,Kovchegov:1996ty}.
Gluon saturation is expected to increase with increasing beam energy.
Therefore, that ridge correlations appear only at high-multiplicity
events at top LHC energies seems to be evidence supporting the
onset of gluon saturation. The way gluon saturation affects particle
production in proton or nucleus collisions is studied via glasma diagrams.
Such calculations indicate that the two-hadron correlation function
calculated from the glasma diagrams explains the systematics of the
ridge signal at the LHC data \cite{Dumitru:2010iy,Dusling:2012iga,Dusling:2012cg,Dusling:2012wy,Dusling:2013oia,Venugopalan:2013cga}. 

Whether the origin of the collectivity observed in experiments is
due to hydrodynamic evolution of the system or due to gluon saturation
is still under discussion. The two-hadron correlations alone are not enough
to settle this dispute. For this purpose, correlations between more
than two hadrons must be measured, and these measurements need to be
compared with the results from hydrodynamics and glasma diagrams separately
\cite{Ozonder:2014sra,Ozonder:2016xqn}. 

Hadron correlation measurements are used to obtain the flow coefficients $v_{m}\{n\text{PC}\}$, where $n\text{PC}$
refers to that the flow coefficient is found from $n$-particle correlations
\cite{Bilandzic:2010jr,Schenke:2017bog,Dusling:2015gta}. Currenty,
the elliptic flow coefficient $v_{2}$ is measured from $n=2,4,6$
and 8 particles in experiments \cite{Adam:2016ows,Aaboud:2017blb,Padula:2016tzw,Belmont:2017lse}.
On the theory side, such coefficients can be calculated from glasma
diagrams. By using glasma diagrams, one can calculate the correlations
of $n$ gluons, and convolving these with the fragmentation functions
results in the hadronic correlation functions that can be compared
with the ones measured by the experiments. In this work, we calculate
five- and six-gluon correlation functions from glasma superdiagrams
towards this goal. 

Another important result of this paper, in addition to the derivation
of these two correlation functions, is the cumulant expansion of the
gluon correlation functions in the gluon saturation limit. An $n$-gluon correlation function is
a cumulant that can be expanded in terms of lower-order cumulants
and $n$th moment {[}see, for example, Eq.~(\ref{k4-standard}){]}.
However, the standard cumulant expansion needs to be modified if one wants to use it to determine the number of glasma diagrams to be calculated.
This has been realized for the first time in the calculation of four-gluon
correlation function in Ref. \cite{Ozonder:2014sra}. Here we derive
the formula that generates the modified cumulant expansion for the $n$-gluon
correlation function obtained from the glasma diagrams. The importance
of this modified cumulant expansion in the context of this work is to find
the number of glasma diagrams to be calculated at a given order and
use it to verify independently that the number of terms in the general
formula that produces the correlation function at $n$th order is
correct.

In the next section, we derive for the first time the formula that
generates the modified cumulant expansion. Then, we review the recipe
developed in Ref. \cite{Ozonder:2016xqn} that yields the $n$-gluon
correlation function. Following that, we will derive the five- and
six-gluon correlation functions and verify by using the modified
cumulant expansion that the number of terms, each of which corresponds
to a connected glasma diagram, in these correlation functions is
correct.

\section{Cumulant Expansion for Rainbow Glasma Diagrams}

The three- and four-gluon azimuthal correlation functions with full
rapidity and transverse momentum dependence have been calculated in
Ref. \cite{Ozonder:2014sra} by using $16$ and $96$ glasma diagrams, respectively.
The observable to be calculated--and later compared to the experimentally
obtained correlation function--for $n$ gluons is the connected azimuthal
correlation function $C_{n}$. Since this function includes only the
connected diagrams, it is a cumulant, not a moment. 

The number of
connected diagrams to be calculated can be determined via the cumulant
expansion. An important realization has been made that the glasma
correlation functions at higher orders, starting with the four-gluon correlation function, 
obeyed the standard cumulant expansion, but one had to modify this expansion if it was to be used to determine the number of glasma diagrams at that order \cite{Ozonder:2014sra}.
Hence, the cumulant expansion should be modified for the rainbow glasma
diagrams when it is to be used to count glasma diagrams. To illustrate via the example of the four-gluon correlation
function, we first write the \textit{standard} cumulant expansion at
the fourth order:
\begin{equation}
\kappa_{4}=\mu_{4}-4\kappa_{3}\kappa_{1}-3\kappa_{2}^{2}-6\kappa_{2}\kappa_{1}^{2}-\kappa_{1}^{4},\label{k4-standard}
\end{equation}
where $\kappa$'s denote the cumulants (connected correlations) and
$\mu_{4}$ denotes the fourth moment which includes all connected
and disconnected glasma diagrams involving four gluons. 
\begin{figure}
\includegraphics[scale=0.8]{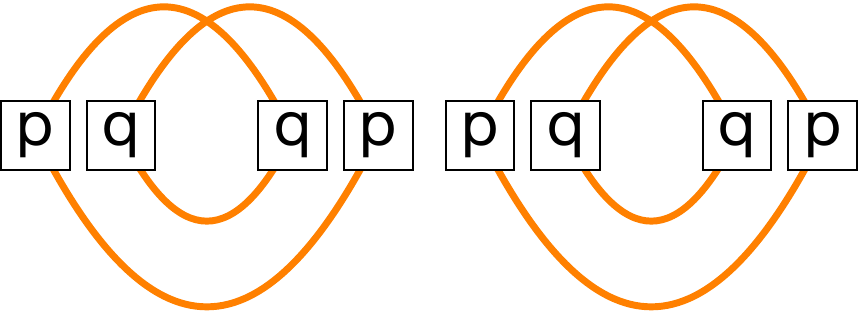}~~~~~~~~~~~~~~~~~~~\includegraphics[scale=0.8]{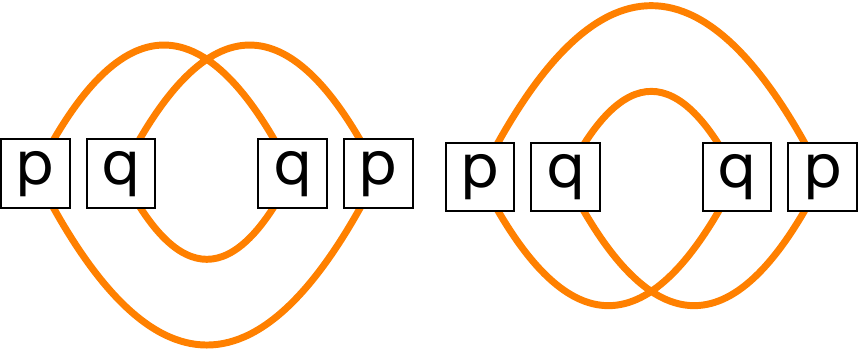}

\caption{Rainbow approximation refers to that either the upper or
lower part of the diagrams includes lines that connect the same momentum
labels ($p\leftrightarrow p$, $q\leftrightarrow q$, etc.). The four
diagrams here separately satisfy this condition. Rainbow approximation,
however, also requires that, when separate diagrams are combined as
in $\kappa_{2}\otimes\kappa_{2}$ in Eq. (\ref{k4-standard}), all
combined diagrams should have their rainbow parts in either their
upper or lower parts. The two disconnected diagrams on the left are
for the term $\kappa_{2}^{\mathrm{low}}\otimes\kappa_{2}^{\mathrm{low}}$,
which is a rainbow diagram altogether. Therefore, this pair of diagrams
is required to be subtracted from the moment $\mu$. However, the
pair on the right ($\kappa_{2}^{\mathrm{low}}\otimes\kappa_{2}^{\mathrm{up}}$)
mixes lower and upper rainbow diagrams. Such terms are already nonexistent
in the fourth moment $\mu_{4}$, so subtracting such diagrams would
lead to wrong counting. }

\label{fig-rainbows}
\end{figure}
 The number of glasma diagrams that each term contains is given by
$\mu_{4}=209$, $\kappa_{3}=16$, $\kappa_{2}=4$ and $\kappa_{1}=1$
\cite{Ozonder:2014sra}. From this, we find $\kappa_{4}=72$, which
is incorrect, since we know that one needs to calculate $96$ connected
rainbow glasma diagrams instead of $72$ as explained in detail in
Ref. \cite{Ozonder:2014sra}. This error occurs since in Eq.~(\ref{k4-standard})
the term $3\kappa_{2}^{2}=3(\kappa_{2}^{\mathrm{up}}+\kappa_{2}^{\mathrm{low}})\otimes(\kappa_{2}^{\mathrm{up}}+\kappa_{2}^{\mathrm{low}})$
mixes the upper and lower glasma diagrams (see Fig. \ref{fig-rainbows}).
The subtraction of the two mixed terms, $\kappa_{2}^{\mathrm{up}}\otimes\kappa_{2}^{\mathrm{low}}$
and $\kappa_{2}^{\mathrm{low}}\otimes\kappa_{2}^{\mathrm{up}}$, gives
rise to wrong counting since in the leading rainbow glasma approximation
such diagrams are already never considered.

In other words, the moment $\mu_{4}$ already does not contain such
correlations, so one should not attempt to subtract the mixed diagrams
($\kappa_{2}^{\mathrm{up}}\otimes\kappa_{2}^{\mathrm{low}}$ and $\kappa_{2}^{\mathrm{low}}\otimes\kappa_{2}^{\mathrm{up}}$)
from $\mu_{4}$. Only the terms $\kappa_{2}^{\mathrm{up}}\otimes\kappa_{2}^{\mathrm{up}}+\kappa_{2}^{\mathrm{low}}\otimes\kappa_{2}^{\mathrm{low}}$
should be subtracted from $\mu_{4}$. Considering that there are the same
number of mixed (up$\otimes$low and low$\otimes$up) and unmixed (up$\otimes$up and low$\otimes$low) glasma diagrams and that we want to keep only the unmixed ones, we can simply substitute $3\kappa_{2}^{2}$ in Eq.~(\ref{k4-standard}) with $3\kappa_{2}^{2}/2$ and
write the \textit{rainbow} cumulant for rainbow glasma diagrams as\footnote{We shall not add any specific identifier index for the \textit{rainbow}
cumulants to distinguish them from the \textit{standard} ones; the
\textit{rainbow} cumulants are recognized by the 2's in the denominators
of some of its terms in the expansion.} 

\begin{equation}
\kappa_{4}=\mu_{4}-4\kappa_{3}\kappa_{1}-3\frac{\kappa_{2}^{2}}{2}-6\kappa_{2}\kappa_{1}^{2}-\kappa_{1}^{4}.
\end{equation}
This gives the correct counting $\kappa_{4}=96$, which matches
the number of connected diagrams calculated explicitly in Ref. \cite{Ozonder:2014sra}.
The \textit{rainbow} cumulant in the fifth order is written as

\begin{equation}
\kappa_{5}=\mu_{5}-5\kappa_{4}\kappa_{1}-10\frac{\kappa_{3}\kappa_{2}}{2}-10\kappa_{3}\kappa_{1}^{2}-15\frac{\kappa_{2}^{2}}{2}\kappa_{1}-10\kappa_{2}\kappa_{1}^{3}-\kappa_{1}^{5},\label{rainbow-k5}
\end{equation}
and at the sixth order it is written as

\begin{align}
\kappa_{6} & =\mu_{6}-6\kappa_{5}\kappa_{1}-15\frac{\kappa_{4}\kappa_{2}}{2}-15\kappa_{4}\kappa_{1}^{2}-10\frac{\kappa_{3}^{2}}{2}-60\frac{\kappa_{3}\kappa_{2}}{2}\kappa_{1}\label{rainbow-k6} \nonumber \\
 & \,\,\,\,\,\,\,\,\,\,\,\,\,\,\,\,\,\,\,\,\,\,-20\kappa_{3}\kappa_{1}^{3}-15\frac{\kappa_{2}^{3}}{4}-45\frac{\kappa_{2}^{2}}{2}\kappa_{1}^{2}-15\kappa_{2}\kappa_{1}^{4}-\kappa_{1}^{6}.
\end{align}

Now we will derive the formula for the \textit{rainbow} cumulants. The
\textit{standard} moment of the order of $n$ in terms of cumulants is given
in terms of the partial Bell polynomials $B_{n,k}$:

\begin{equation}
\mu_{n}=\sum_{k=1}^{n}B_{n,k}(\kappa_{1},\kappa_{2},\ldots,\kappa_{n-k+1}),\label{standard-moment}
\end{equation}
and \textit{standard} $\kappa_{n}$ is found by solving this equation
for $\kappa_{n}$. From the discussions above, we can write the expression
for the \textit{rainbow }moment as\footnote{The Mathematica code for both the \textit{standard} and \textit{rainbow}
cumulants are given in the Appendix.}

\begin{equation}
\mu_{n}=-\kappa_{1}^{n}+2\sum_{k=1}^{n}B_{n,k}\left(\kappa_{1},\frac{\kappa_{2}}{2},\ldots,\frac{\kappa_{n-k+1}}{2}\right),\label{rainbow-moment}
\end{equation}
and the \textit{rainbow} cumulant $\kappa_{n}$ is found by solving
this equation for $\kappa_{n}$. The formula in Eq. (\ref{rainbow-moment})
is the first result of this paper.\footnote{To emphasize, the non-
standard, \textit{rainbow} cumulant expansion is developed here for correctly counting the glasma diagrams. The correlation functions $C_n$ still obey the standard cumulant expansion. This can be understood from the discussion following Eq. (46) in Ref. \cite{Ozonder:2014sra}. The Eq. (46) therein is clearly in the form of standard cumulant expansion at the order $n=4$.}
After we derive the five- and six-gluon correlation functions later
in this work, we will use the \textit{rainbow} cumulant expansion
in Eq. (\ref{rainbow-moment}) to check if the five- and six-gluon
correlation functions include the correct number of terms, where each
term corresponds to one connected diagram.

\section{Formula of $n$-Gluon Correlation Function}

In Ref. \cite{Ozonder:2016xqn}, we derived the formula for the $n$-gluon
correlation function $C_{n}$ with full momentum and rapidity dependence
by using the glasma superdiagrams we developed. Here we quote the
formulas that we will use in the next section to derive five- and
six-gluon correlation functions.

The $n$-gluon correlation function is given by

\begin{equation}
C_{n}=\frac{\alpha_{s}^{n}N_{c}^{n}S_{\perp}}{\pi^{4n}(N_{c}^{2}-1)^{2n-1}}\left(\prod_{i=1}^{n}\frac{1}{\boldsymbol{p}_{\perp i}^{2}}\right)\int\frac{d^{2}\boldsymbol{k}_{\perp}}{(2\pi)^{2}}\left({\cal N}_{1}+{\cal N}_{2}\right).\label{C_n}
\end{equation}
Here $\alpha_{s}$ is the QCD strong coupling constant, $N_{c}=3$
is the gluon color factor, $S_{\perp}$ is the transverse area of
overlap during the collision between the target and projectile, and $p_{\perp i}$
are the transverse momentum variables of the gluons produced. ${\cal N}_{1,2}$,
which include the unintegrated gluon distribution (UGD) functions
$\Phi_{A,p}(\boldsymbol{p}_{\perp})$, are given 
by\footnote{The formulas for ${\cal N}_{1,2}$ given in Ref. \cite{Ozonder:2016xqn} included a typo and missed the prefactor $f_n$, and the prefactor in the second brackets mistakenly read $2^h$ instead of $2h$. These mistakes in Ref. \cite{Ozonder:2016xqn} originated from the miscalculation of the $\kappa_5$ therein.}

\begin{align}
{\cal N}_{1} & =f_{n}\Phi_{1,p_{1}}^{2}(\boldsymbol{k}_{\perp})\left[\prod_{j=1}^{n-3}\Phi_{1,p_{j+1}}(\boldsymbol{k}_{\perp})\right]\left[\sum_{h=1}^{n-2}2h\Phi_{1,p_{h+1}}(\boldsymbol{k}_{\perp})\right]\Phi_{2,p_{1}}(\boldsymbol{p}_{\perp1}-\boldsymbol{k}_{\perp})\,{\cal N}_{A_{2}},\label{N1}\\
{\cal N}_{2} & =f_{n}\Phi_{2,p_{n}}^{2}(\boldsymbol{k}_{\perp})\left[\prod_{j=1}^{n-3}\Phi_{2,p_{n-j}}(\boldsymbol{k}_{\perp})\right]\left[\sum_{h=1}^{n-2}2h\Phi_{2,p_{n-h}}(\boldsymbol{k}_{\perp})\right]\Phi_{1,p_{1}}(\boldsymbol{p}_{\perp1}-\boldsymbol{k}_{\perp})\,{\cal N}_{A_{1}},\label{N2}
\end{align}
where
\begin{equation}
{\cal N}_{A_{1}(A_{2})}=\prod_{m=2}^{n}\left[\Phi_{1(2),p_{m}}(\boldsymbol{p}_{\perp m}-\boldsymbol{k}_{\perp})+\Phi_{1(2),p_{m}}(\boldsymbol{p}_{\perp m}+\boldsymbol{k}_{\perp})\right].\label{NA1NA2}
\end{equation}
The indices of $\Phi_{A,p}(\boldsymbol{p}_{\perp})$ are as follows.
$A$ stands for the target or projectile index ($A=1,2$), $p$ subscript
is the rapidity variable of the gluon, and $\boldsymbol{p}_{\perp}$
is the transverse momentum variable of the gluon.

The coefficient $f_{n}$ is given by

\begin{equation}
f_{n}=\begin{cases}
\begin{array}{c}
1\\
(n-3)!
\end{array} & \begin{array}{c}
\text{if }n<5,\\
\text{if }n\geq5,
\end{array}\end{cases}
\end{equation}
where $n$ here is the same $n$ as in $C_{n}$, i.e., number of gluons.

It is important to note that the rapidity indices and that which UGD
appears with which prefactor in the formulas above are nontrivial.
The former is found by means of superdiagrams, and the latter is found
by means of the \textit{rainbow }cumulant expansion. In this work,
we will show only the relevant superdiagrams for $C_{5}$ and $C_{6}$,
but we will not explain how they are drawn, for which we refer the
interested reader to our earlier work \cite{Ozonder:2016xqn}.

\section{Five-gluon Azimuthal Correlation Function}

\begin{figure}
\includegraphics[scale=0.3]{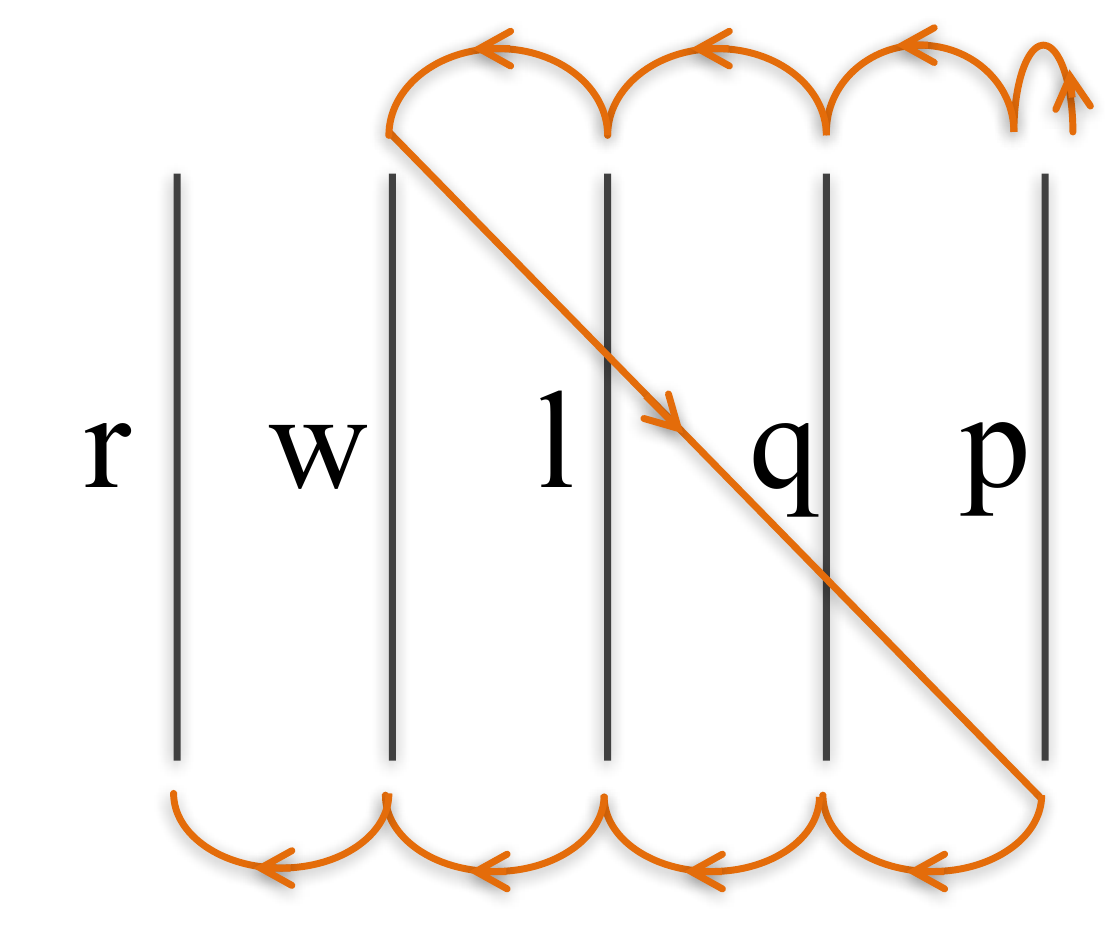}~~~~\includegraphics[scale=0.3]{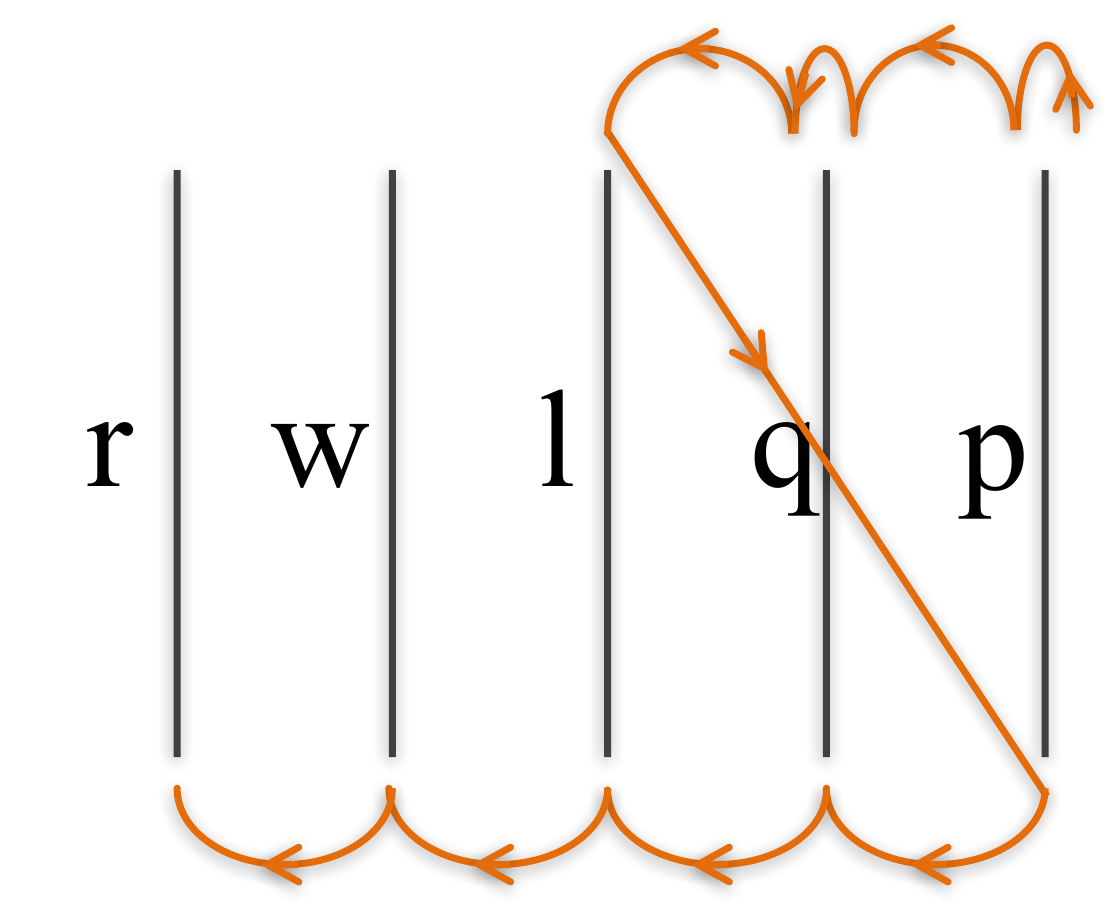}~~~~\includegraphics[scale=0.3]{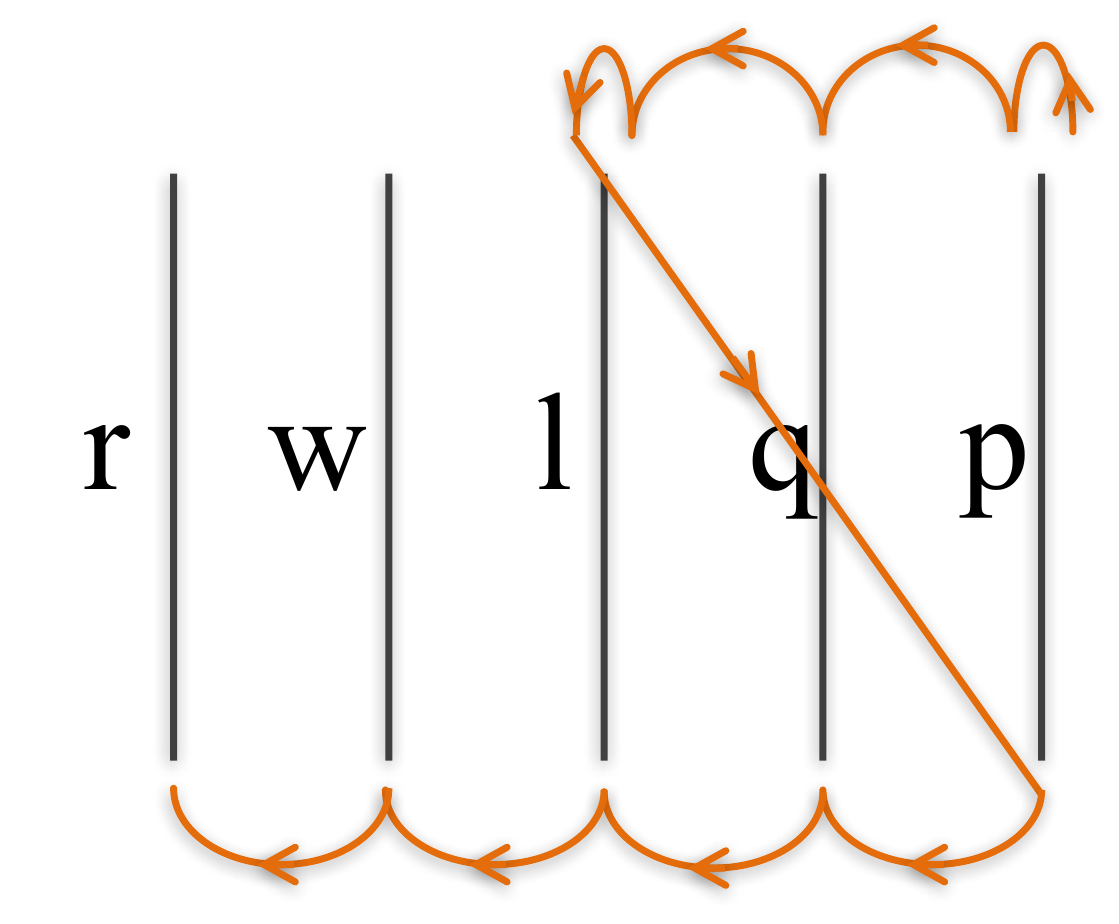}
\caption{Superdiagrams for $C_{5}$. These three superdiagrams
give rise to ${\cal N}_{1}^{(5)}$ in Eq. (\ref{N1-5}), particularly
the rapidity indices of the unintegrated distribution functions $\Phi$.
There are three other superdiagrams that are diagonally mirror images
of these; i.e., they run starting from the bottom left, and end at the top right. 
Those diagrams give ${\cal N}_{2}^{(5)}$ in Eq. (\ref{N2-5}).}
\label{fig-C5}
\end{figure}

The five-gluon correlation function $C_{5}$ can be written by using
the formulas given in Eqs.~(\ref{C_n})-(\ref{NA1NA2})
as follows:

\begin{equation}
C_{5}(\boldsymbol{p},\boldsymbol{q},\boldsymbol{l},\boldsymbol{w},\boldsymbol{r})=\frac{\alpha_{s}^{5}N_{c}^{5}S_{\perp}}{\pi^{20}(N_{c}^{2}-1)^{9}}\frac{1}{\boldsymbol{p}_{\perp}^{2}\boldsymbol{q}_{\perp}^{2}\boldsymbol{l}_{\perp}^{2}\boldsymbol{w}_{\perp}^{2}\boldsymbol{r}_{\perp}^{2}}\int\frac{d^{2}\boldsymbol{k}_{\perp}}{(2\pi)^{2}}\left({\cal N}_{1}^{(5)}+{\cal N}_{2}^{(5)}\right),\label{C5}
\end{equation}
where

\begin{align}
{\cal N}_{1}^{(5)} & =2\,\Phi_{1,p}^{2}(\boldsymbol{k}_{\perp})\Phi_{1,q}(\boldsymbol{k}_{\perp})\Phi_{1,l}(\boldsymbol{k}_{\perp})\left[2\Phi_{1,q}(\boldsymbol{k}_{\perp})+4\Phi_{1,l}(\boldsymbol{k}_{\perp})+6\Phi_{1,w}(\boldsymbol{k}_{\perp})\right]\Phi_{2,p}(\boldsymbol{p}_{\perp}-\boldsymbol{k}_{\perp})\,{\cal N}_{A_{2}}^{(5)},\label{N1-5}\\
{\cal N}_{2}^{(5)} & =2\,\Phi_{2,r}^{2}(\boldsymbol{k}_{\perp})\Phi_{2,w}(\boldsymbol{k}_{\perp})\Phi_{2,l}(\boldsymbol{k}_{\perp})\left[2\Phi_{2,w}(\boldsymbol{k}_{\perp})+4\Phi_{2,l}(\boldsymbol{k}_{\perp})+6\Phi_{2,q}(\boldsymbol{k}_{\perp})\right]\Phi_{1,p}(\boldsymbol{p}_{\perp}-\boldsymbol{k}_{\perp})\,{\cal N}_{A_{1}}^{(5)},\label{N2-5}
\end{align}
and
\begin{align}
{\cal N}_{A_{1}(A_{2})}^{(5)} & =\left[\Phi_{1(2),q}(\boldsymbol{q}_{\perp}-\boldsymbol{k}_{\perp})+\Phi_{1(2),q}(\boldsymbol{q}_{\perp}+\boldsymbol{k}_{\perp})\right]\left[\Phi_{1(2),l}(\boldsymbol{l}_{\perp}-\boldsymbol{k}_{\perp})+\Phi_{1(2),l}(\boldsymbol{l}_{\perp}+\boldsymbol{k}_{\perp})\right]\nonumber \\
 & \times\left[\Phi_{1(2),w}(\boldsymbol{w}_{\perp}-\boldsymbol{k}_{\perp})+\Phi_{1(2),w}(\boldsymbol{w}_{\perp}+\boldsymbol{k}_{\perp})\right]\left[\Phi_{1(2),r}(\boldsymbol{r}_{\perp}-\boldsymbol{k}_{\perp})+\Phi_{1(2),r}(\boldsymbol{r}_{\perp}+\boldsymbol{k}_{\perp})\right]\label{NA-5}
\end{align}

The $n$th level moment $\mu_{n}$ includes both the connected and
disconnected glasma diagrams for $n$-gluon correlation functions,
and it is given by $\mu_{n}=2(2n-1)!!-1$. Using Eq. (\ref{rainbow-k5})
and that $\mu_{5}=1889$, $\kappa_{1}=1$, $\kappa_{2}=4$, $\kappa_{3}=16$,
and $\kappa_{4}=96$, one finds $\kappa_{5}=768$. So, one needs to
consider 768 connected glasma diagrams to calculate $C_{5}$. Equation~(\ref{NA-5}) has $2^4$ terms, so Eq.~(\ref{N1-5}) contains $2\times(2+4+6)\times2^4=384$ terms. Similarly, Eq.~(\ref{N2-5}) contains 384 terms as well. So, $C_{5}$ in Eq.~(\ref{C5}) includes exactly 768 terms, which matches with the number that we find from the rainbow cumulant expansion.

It would be practically impossible to calculate such number of diagrams
separately without the superdiagrams, where one needs only three superdiagrams
for $C_{5}$ (see Fig. \ref{fig-C5}). For $C_{6}$, one needs to
calculate 7680 connected glasma diagrams or only four superdiagrams.

\section{Six-gluon Azimuthal Correlation Function}

\begin{figure}
\includegraphics[scale=0.3]{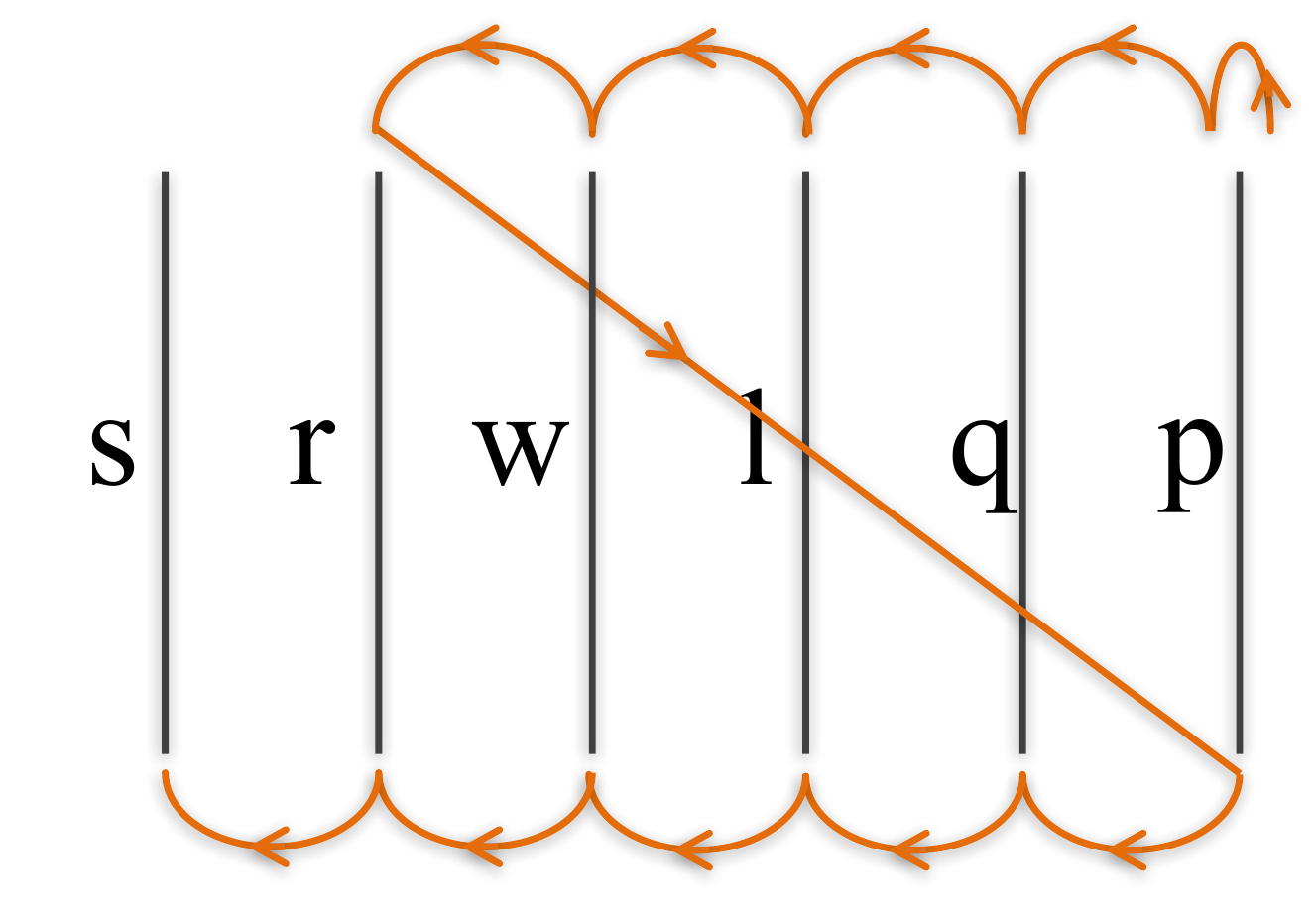}~~~~\includegraphics[scale=0.3]{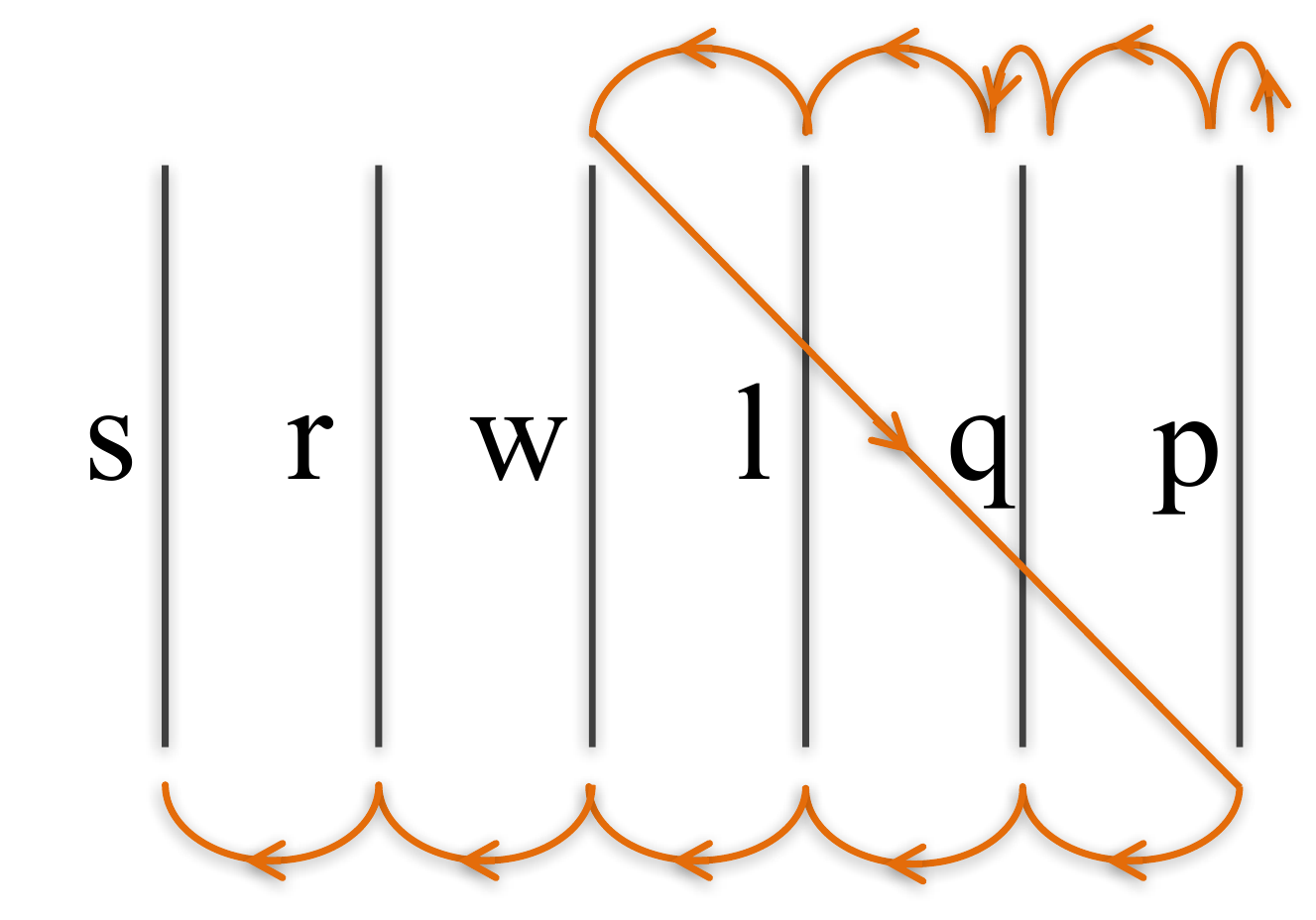}~~~~\includegraphics[scale=0.3]{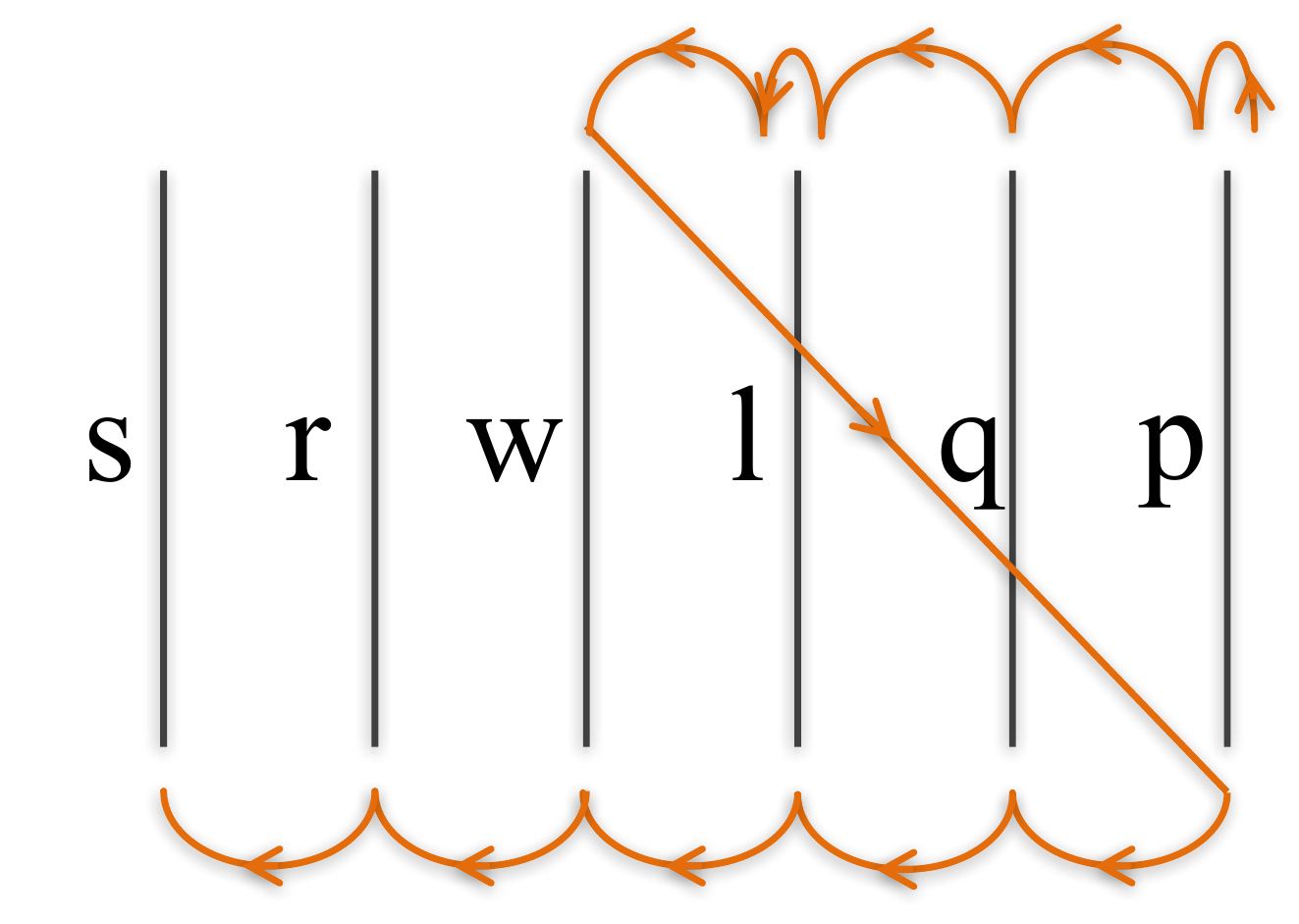}~~~~\includegraphics[scale=0.3]{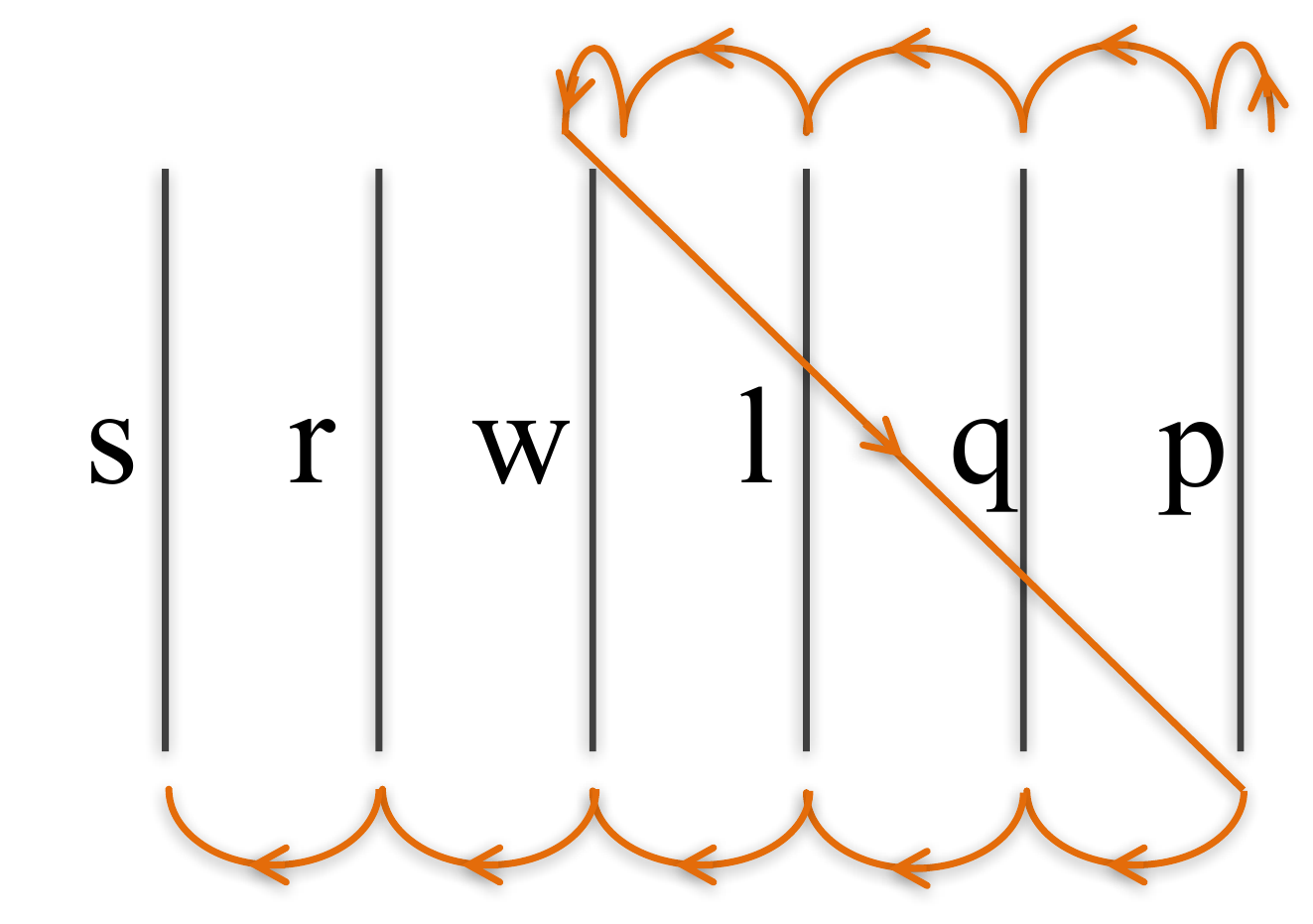}
\caption{Superdiagrams for $C_{6}$. These four superdiagrams
give rise to ${\cal N}_{1}^{(6)}$ in Eq. (\ref{N1-6}). There are
four other superdiagrams that are diagonally mirror images of these;
i.e., they run starting from the bottom left and end at the top right. 
Those diagrams
give ${\cal N}_{2}^{(6)}$ in Eq. (\ref{N2-6}).}
\label{fig-C6}
\end{figure}

The six-gluon correlation function $C_{6}$ can be written by using
the formulas given in Eqs.~(\ref{C_n})-(\ref{NA1NA2})
as follows:

\begin{equation}
C_{6}(\boldsymbol{p},\boldsymbol{q},\boldsymbol{l},\boldsymbol{w},\boldsymbol{r},\boldsymbol{s})=\frac{\alpha_{s}^{6}N_{c}^{6}S_{\perp}}{\pi^{24}(N_{c}^{2}-1)^{11}}\frac{1}{\boldsymbol{p}_{\perp}^{2}\boldsymbol{q}_{\perp}^{2}\boldsymbol{l}_{\perp}^{2}\boldsymbol{w}_{\perp}^{2}\boldsymbol{r}_{\perp}^{2}\boldsymbol{s}_{\perp}^{2}}\int\frac{d^{2}\boldsymbol{k}_{\perp}}{(2\pi)^{2}}\left({\cal N}_{1}^{(6)}+{\cal N}_{2}^{(6)}\right),\label{C6}
\end{equation}
where 

\begin{align}
{\cal N}_{1}^{(6)} & =6\,\Phi_{1,p}^{2}(\boldsymbol{k}_{\perp})\Phi_{1,q}(\boldsymbol{k}_{\perp})\Phi_{1,l}(\boldsymbol{k}_{\perp})\Phi_{1,w}(\boldsymbol{k}_{\perp})\left[2\Phi_{1,q}(\boldsymbol{k}_{\perp})+4\Phi_{1,l}(\boldsymbol{k}_{\perp})+6\Phi_{1,w}(\boldsymbol{k}_{\perp})+8\Phi_{1,r}(\boldsymbol{k}_{\perp})\right]\nonumber \\
 & \,\,\,\,\,\,\,\,\,\,\,\,\,\,\,\times\Phi_{2,p}(\boldsymbol{p}_{\perp}-\boldsymbol{k}_{\perp})\,{\cal N}_{A_{2}}^{(6)},\label{N1-6}\\
{\cal N}_{2}^{(6)} & =6\,\Phi_{2,s}^{2}(\boldsymbol{k}_{\perp})\Phi_{2,r}(\boldsymbol{k}_{\perp})\Phi_{2,w}(\boldsymbol{k}_{\perp})\Phi_{2,l}(\boldsymbol{k}_{\perp})\left[2\Phi_{2,r}(\boldsymbol{k}_{\perp})+4\Phi_{2,w}(\boldsymbol{k}_{\perp})+6\Phi_{2,l}(\boldsymbol{k}_{\perp})+8\Phi_{1,q}(\boldsymbol{k}_{\perp})\right]\nonumber \\
 & \,\,\,\,\,\,\,\,\,\,\,\,\,\,\,\times\Phi_{1,p}(\boldsymbol{p}_{\perp}-\boldsymbol{k}_{\perp})\,{\cal N}_{A_{1}}^{(5)},\label{N2-6}
\end{align}
and
\begin{align}
{\cal N}_{A_{1}(A_{2})}^{(6)} & =\left[\Phi_{1(2),q}(\boldsymbol{q}_{\perp}-\boldsymbol{k}_{\perp})+\Phi_{1(2),q}(\boldsymbol{q}_{\perp}+\boldsymbol{k}_{\perp})\right]\left[\Phi_{1(2),l}(\boldsymbol{l}_{\perp}-\boldsymbol{k}_{\perp})+\Phi_{1(2),l}(\boldsymbol{l}_{\perp}+\boldsymbol{k}_{\perp})\right]\nonumber \\
 & \times\left[\Phi_{1(2),w}(\boldsymbol{w}_{\perp}-\boldsymbol{k}_{\perp})+\Phi_{1(2),w}(\boldsymbol{w}_{\perp}+\boldsymbol{k}_{\perp})\right]\left[\Phi_{1(2),r}(\boldsymbol{r}_{\perp}-\boldsymbol{k}_{\perp})+\Phi_{1(2),r}(\boldsymbol{r}_{\perp}+\boldsymbol{k}_{\perp})\right]\nonumber \\
 & \times\left[\Phi_{1(2),s}(\boldsymbol{s}_{\perp}-\boldsymbol{k}_{\perp})+\Phi_{1(2),s}(\boldsymbol{s}_{\perp}+\boldsymbol{k}_{\perp})\right].
\end{align}

Using Eq. (\ref{rainbow-k6}) and that $\mu_{6}=2(2\times6-1)!!-1=20789$,
$\kappa_{1}=1$, $\kappa_{2}=4$, $\kappa_{3}=16$, $\kappa_{4}=96$,
and $\kappa_{5}=768$, one finds $\kappa_{6}=7680$. $C_{6}$ in Eq.
(\ref{C6}) includes exactly 7680 terms, so this verifies that Eq.
(\ref{C6}) is correct. The superdiagrams needed for calculating $C_6$ are given in Fig. \ref{fig-C6}.

\section{Conclusion}

In this paper, we derived the modified cumulant expansion that was to be used for counting glasma diagrams correctly.
This expansion is
essential in deriving the correlation functions and verifying that
they are found from the correct glasma diagrams. We also derived the
five- and six-gluon correlation functions. The six-gluon correlation function,
particularly, can be used to calculate the elliptic flow $v_{2}\{6\}$,
and then it can be compared with the measurements. Another use of
the correlation functions we calculated could be studying correlations
between flow coefficients \cite{ALICE:2016kpq} as well as obtaining
other observables from the ratios of the cumulants ($C_{n}$'s) \cite{Bzdak:2017ltv}.
These studies are underway.

\section*{Appendix: Codes for the cumulants in Mathematica}

The Mathematica function which gives the \textit{standard} cumulant
expansion of $\kappa_{n}$ is {[}see Eq. (\ref{standard-moment}){]}

\begin{equation}
\kappa[n\_]:=\kappa_{n}+\mu_{n}-\sum_{k=1}^{n}\text{BellY}\left[n,k,\text{Table}\left[\kappa_{m},\{m,n\}\right]\right],
\end{equation}
and that which gives the \textit{rainbow} cumulant expansion of $\kappa_{n}$
is {[}see Eq. (\ref{rainbow-moment}){]} 

\begin{equation}
\kappa[n\_]:=\kappa_{1}^{n}+\kappa_{n}+\mu_{n}-2\sum_{k=1}^{n}\text{BellY}\left[n,k,\text{Join}\left[\{\kappa_{1}\},\text{Table}\left[\frac{\kappa_{m}}{2},\{m,2,n\}\right]\right]\right].
\end{equation}

\section{Acknowledgement}

This work is supported by the grant TUBITAK BIDEB 2232-117C008.

\bibliographystyle{apsrev4-1}
\bibliography{glasma-cumulants}

\end{document}